\documentclass[oneside,a4paper,10pt]{llncs}

\usepackage[english]{babel}
\usepackage{amssymb}
\usepackage{graphicx}
\usepackage{units}
\usepackage[usenames,dvipsnames]{xcolor}
\usepackage{fixltx2e}
\usepackage{booktabs}
\usepackage{setspace}
\usepackage{listings}
\usepackage{footmisc}
\usepackage{wrapfig}
\usepackage{paralist}
\usepackage{caption}
\usepackage{subcaption}
\usepackage{url}
\usepackage{color}

\definecolor{TUMblue}{RGB}   {  0, 101, 189} 
\definecolor{Pantone540}{RGB}{  0,  51,  89}
\definecolor{Pantone301}{RGB}{  0,  82, 147}
\definecolor{Pantone285}{RGB}{  0, 115, 207}
\definecolor{Pantone542}{RGB}{100, 160, 200}
\definecolor{Pantone283}{RGB}{152, 198, 234}
\definecolor{TUMdarkgray}{RGB}{ 88, 88, 90}
\definecolor{TUMgray}{RGB}{ 156, 157, 159}
\definecolor{TUMlightgray}{RGB}{ 217, 218, 219}
\definecolor{TUMgreen}{RGB}{162, 173, 0} 
\definecolor{TUMorange}{RGB}{227, 114, 34} 
\definecolor{TUMelfenbein}{RGB}{218, 215, 203} 
\definecolor{TUMpyellow}{RGB}{255, 180, 0}
\definecolor{TUMporange}{RGB}{255, 128, 0}
\definecolor{TUMpred}{RGB}{229, 52, 24}
\definecolor{TUMpdarkred}{RGB}{202, 33, 63}
\definecolor{TUMpblue}{RGB}{0, 153, 255}
\definecolor{TUMplightblue}{RGB}{65, 190, 255}
\definecolor{TUMpgreen}{RGB}{145, 172, 107}
\definecolor{TUMplightgreen}{RGB}{181, 202, 130}
\definecolor{TUMpbluethemelower}{RGB}{0, 82, 147} 
\definecolor{TUMpbluethemeupper}{RGB}{0, 40, 72}

\usepackage[a4paper,top=1.62in,bottom=2.5in,left=1.87in,right=1.59in]{geometry}

\let\subparagraph\paragraph
\usepackage{titlesec}
\usepackage{chngcntr}
\usepackage{hyperref}
\usepackage{siunitx}
\usepackage{tikz}
\usetikzlibrary{shapes,arrows,shadows,fit}

\definecolor{shadecolor}{gray}{.95}                                                           

\begin{document}

\title{Development and Verification of a Flight Stack\\for a High-Altitude Glider in Ada/SPARK~2014\thanks{The source code for this project is available at \url{github.com/tum-ei-rcs/StratoX}. This is a pre-print. The final publication will be available at Springer via \url{http://dx.doi.org/TODO} in \emph{Computer Safety, Reliability and Security, 36th International Conference SAFECOMP~2017}, S. Tonetta and E. Schoitsch and F. Bitsch (Eds.), Trento, Italy, 2017.}}
\titlerunning{A SPARK~2014 Flight Stack for a High-Altitude Glider}  
%
\author{\vspace*{-1em}Martin Becker \and Emanuel Regnath \and Samarjit Chakraborty}
\authorrunning{Becker et al.} 
%
\tocauthor{Martin Becker, Emanuel Regnath, and Samarjit Chakraborty}
\institute{Chair of Real-Time Computer Systems, Technical University of Munich\\80333 Munich, Germany\\
}

\maketitle              

\begin{abstract}\vspace*{-1em}
%
  SPARK~2014 is a modern programming language and a new state-of-the-art
  tool set for development and verification of high-integrity
  software. In this paper, we explore the capabilities and limitations
  of its latest version in the context of building a flight stack for
  a high-altitude unmanned glider. Towards that, we deliberately
  applied static analysis early and continuously during
  implementation, to give verification the possibility to steer
  the software design. 
  In this process we have identified several limitations and pitfalls
  of software design and verification in SPARK, for which we give
  workarounds and protective actions to avoid them. Finally, we give
  design recommendations that have proven effective for verification, and
  summarize our experiences with this new language.
\keywords{Ada/SPARK, formal verification, limitations, rules}
\end{abstract}
\section{Introduction}
\label{sec:intro}




The system under consideration is a novel kind of weather balloon
which is actively controlled, and thus requires verification to ensure
it is working properly in public airspace.  As any normal weather
balloon, the system climbs up to the stratosphere (beyond an altitude of $\SI{10}{\kilo\meter}$),
while logging weather data such as temperature, pressure,
NO\textsubscript{2}-levels and so on. Eventually the balloon bursts,
and the sensors would be falling back to the ground with a parachute,
drifting away with prevailing wind conditions. However, our system is
different from this point onwards: the sensors are placed in a
light-weight glider aircraft which is attached to the balloon. At a
defined target altitude, the glider separates itself from the balloon,
stabilizes its attitude and performs a controlled descent back to the
take-off location, thus, bringing the sensors back home. In this paper
we focus on the development and formal verification of the glider's
onboard software.



The requirements for such a system are challenging already because of the
extreme environmental conditions; temperatures range from $\SI{+30}{\celsius}$
down to $\SI{-50}{\celsius}$, winds may exceed $\SI{100}{kph}$, and GPS devices may
yield vastly different output in those altitudes due to decreasing precision and the
wind conditions. The combination of
those extreme values is likely to trigger corner cases 
in the software, and thus should be covered by means of extensive testing or
by analysis. 

 We use this opportunity of a safety-critical, yet hardly testable
 system to explore the new state-of-the art verification tools of
 Ada/SPARK~2014~\cite{GNATprove}, especially to identify limitations,
 pitfalls and applicability in practice.  To
 experience this new SPARK release to its full extent, we applied a
 \emph{co-verification} approach. That is, we did not perform
 verification on a finished product, but instead in parallel to the
 software development (the specific strategy is not of relevance for this paper, but only the effect that this enabled us to identify 
 code features that pose challenges in verification, and find workarounds for them). The implementation could therefore be shaped by
 verification needs. Moreover, since the high-altitude glider was a
 research project, we allowed ourselves to modify the initial
 software design to ease verification when needed.




\section{Verification in Ada/SPARK}
\label{sec:verification-spark}

SPARK~2014 is a major redesign of the original SPARK language, which was intended for formal verification. SPARK~2014 now adopts Ada~2012 syntax, and covers a large subset of Ada. As a result, the GNAT Ada compiler can build
an executable from SPARK~2014 source code, and even compile a program which mixes both languages. Compared to
Ada, the most important exclusions are
pointers (called \emph{access}), aliasing and allocators, as well as a
ban of exception handling.
As a consequence, SPARK programs first and foremost must be shown to
be free of run-time exceptions (called \emph{AoRTE} - absence of
run-time errors), which constitutes the main verification task.

The SPARK language -- for the rest of this paper we refer to SPARK~2014 simply as SPARK -- is built
on functional contracts and data flow contracts. Subprograms (procedures and
functions) can be annotated with pre- and postconditions, as well as with
data dependencies. GNATprove, the (only) static analyzer for SPARK~2014, aims to prove subprograms in a
modular way, by analyzing each of them individually.
The effects of callees are summarized by their
post-condition when the calling subprogram is analyzed, and the
precondition of the callee is imposing a proof obligation on the
caller, i.e., the need to verify that the caller respects the callee's
precondition. Further proof obligations arise from each
language-defined check that is executed on the target, such as
overflow checks, index checks, and so on. If all proofs are successful, 
then the program is working according to its contracts 
and no exceptions will be raised during execution, i.e., AoRTE is established. 

Internally, GNATprove~\cite{GNATprove} builds on the
Why3 platform~\cite{why3}, which performs deductive verification on
the proof obligations to generate verification conditions (VCs), and
then passes them to a theorem solver of user's choice, e.g., cvc4, alt-ergo or z3. Note that there exists also a tool for abstract interpretation, which is, however, not discussed here.

\subsection{The GNAT Dimensionality Checking System}
We also want to introduce a feature that is not part of the SPARK
language itself, but an implementation-defined extension of the GNAT compiler, and thus available for SPARK programs. 
%
%
Since Ada 2012, the GNAT compiler offers a dimension system for numeric types through implementation-defined aspects~\cite{Schonberg:2012:ISD:2402676.2402692}. 
The dimension system can consist of up to seven base dimensions, 
and physical quantities are declared as subtypes, annotated with the exponents of each dimension.
Expressions using such variables are statically analyzed by the compiler for their dimensional consistency.
Furthermore, the dimensioned variables contribute to readability and documentation of the code. Inconsistencies such as the following are found (dividend and divisor are switched in the calculation of rate):

\vspace{-1mm}\begin{lstlisting}[name=units]
angle : Angle_Type := 20.0 * Degree;         
dt    : Time_Type  := 100.0 * Milli * Second;
rate  : Angular_Velocity_Type := dt / angle; -- compiler error
\end{lstlisting}\vspace{-1mm}
Note that scaling prefixes like \lstinline$Milli$ can be used, and
that common conversions, such as between Degree and Radian in line 1, can be governed in a similar way.

%

In our project, we specified a unit system with the dimensions \emph{length}, \emph{mass}, \emph{time}, \emph{temperature}, \emph{current}, and \emph{angle}. 
Adding angle as dimension provides better protection against assignments of dimensionless types, as proposed in~\cite{Xiang2015}.



\section{Initial System Design \& Verification Goals}
\label{sec:context}

\textbf{Target Hardware.} We have chosen the ``Pixhawk''
autopilot~\cite{pixhawk2011}. It comprises two ARM processors; one
Cortex-M4F (STM32F427) acting as flight control computer, and one
Cortex-M3 co-processor handling the servo outputs. We implemented our
flight stack on the Cortex-M4 from the ground up, thus completely
replacing the original PX4/NuttX firmware that is installed when
shipped. 


\textbf{Board Support, Hardware Abstraction Layer \& Run-Time System.}
We are hiding the specific target from the application
layer by means of a board support package (not to be confused with an Ada package). This package contains an
hardware abstraction layer (HAL) and a run-time system (RTS). The RTS
is implementing basic functionality such
as tasking and 
memory management. 
The HAL is our extension of AdaCore's Drivers
Library~\cite{AdaDriverLib}, and the RTS is our port of the Ada RTS for the
 STM32F409 target.  Specifically, we have ported the
Ravenscar Small Footprint variant~\cite{Ravenscar},   
which restricts Ada's and SPARK's tasking facilities to a deterministic and
analyzable subset, but meanwhile forbids exception handling, which anyway is not permitted in SPARK.


\textbf{Separating Tasks by Criticality\label{sec:separ-tasks-crit}}
has been one goal, since multi-threading is supported in
SPARK. In particular, \begin{inparaenum}
\item termination of low-critical tasks must not cause termination of
  high-critical tasks,
\item higher-criticality tasks must not be blocked by lower-critical tasks and,
\item adverse effects such as deadlocks,
priority inversion and race condition must not occur. 
\end{inparaenum}
We partitioned our glider software into two tasks (further concurrency arises
from interrupt service routines):
\begin{enumerate}
\item The \emph{Flight-Critical Task} includes all execution flows
  required to keep the glider in a controlled and
  navigating flight, thus including sensor reads and actuator
  writes. It is time-critical for control reasons. High-criticality.
\item The \emph{Mission-Critical Task} includes all execution flows that
  are of relevance for recording and logging of weather data to an SD
  card. Low-priority task, only allowed to run when the
  flight-critical task is idle. Low-criticality.
\end{enumerate}
The latter task requires localization data from the former one, to
annotate the recorded weather data before writing it to the SD card.
Additionally, it takes over the role of a flight logger, saving data
from the flight-critical task that might be of interest for a
post-flight analysis. The interface between these two tasks would
therefore be a protected object with a message queue that must be
able to hold different types of messages. 

\textbf{Verification Goals.} First and foremost, AoRTE shall be
established for all SPARK parts, since exceptions would result in task
termination. Additionally, the application shall make use of as many
contracts and checks as possible, and perform all of its computations using
dimension-checked types. Last but not least, a few functional high-level
requirements related to the homing functionality have been encoded in
contracts.
Overall, the focus of verification was the application, not the
BSP. The BSP has been written in SPARK only as far as necessary to 
support proofs in the application. The rationale was that the RTS was assumed to
be well tested, and the HAL was expected to be hardly verifiable
due to direct hardware access involving pointers and restricted types.

\section{Problems and Workarounds}
\label{sec:workarounds}
In this section, we describe the perils and difficulties that we
identified during verification of SPARK programs. We use the following nomenclature:
\begin{itemize}
\item \textbf{False Positive} denotes a failing check (failed VC) in static analysis 
  which would not fail in any execution on the target, i.e., a false alarm.
\item \textbf{False Negative} denotes a successful check (discharged VC) in static
  analysis which 
  would fail in at least one execution on the target, i.e., a missed
  failure.
\end{itemize}

\subsection{How to Miss Errors}
There are a few situations in which static analysis can miss run-time
exceptions, which in a SPARK program inevitably ends in abnormal
program termination. Before we show these unwanted situations, we have to
point out one important property of a deductive verification approach: Proofs
build on each other. Consider the following example (results of static
analysis given in comments):
\begin{lstlisting}[name=proofdep]
  a := X / Z; -- medium: division check might fail
  b := Y / Z; -- info: division check proved
\end{lstlisting}  
The analyzer reports that the check in line 2 cannot fail, although it
suffers from the same defect as line 1. However, when the run-time
check at line 1 fails, then line 2 cannot be reached with the
offending value of \lstinline$Z$, therefore line 2 is
not a False Negative, unless exceptions have been wrongfully disabled.

\textbf{Mistake 1: Suppressing False Positives.} When a developer
comes to the conclusion that the analyzer has generated a False
Positive (e.g., due to insufficient knowledge on something that is
relevant for a proof), then it might be justified to suppress the
failing property. However, we experienced cases where this has generated
False Negatives which where hiding (critical) failures.  Consider the
following code related to 
the GPS: 
\begin{lstlisting}[name=missedbug,escapechar=\$]
function toInt32 (b : Byte_Array) return Int_32 with Pre => b'Length = 4;
procedure Read_From_Device (d : out Byte_Array) is begin
  d := (others => 0); -- False Positive
  pragma Annotate (GNATprove, False_Positive, "length check might fail", ...);
end Read_From_Device;

procedure Poll_GPS is
  buf    : Byte_Array(0..91) := (others => 0);
  alt_mm : Int_32;
begin
  Read_From_Device (buf);
  alt_mm := toInt32(buf(60..64)); -- False Negative, guaranteed exception
end Poll_GPS;
\end{lstlisting}
Static analysis found that the initialization of the array \lstinline$d$
in line 3 could fail, but this is not possible in this context, and
thus a False Positive\footnote{This particular case has been fixed in recent versions of GNATprove.}. The developer
was therefore suppressing this warning with an annotation
pragma. However, because proofs build on each other, a severe defect
in line 12 was missed. The array slice has an off-by-one error which
\emph{guaranteed} failing the precondition check of
\lstinline$toInt32$.  The reason for this False Negative is that
everything after the initialization of \lstinline$d$ became virtually
\emph{unreachable} and that all following VCs consequently have been discharged.  
In general, a False Positive may exclude some or all execution paths
for its following statements, and thus
hide (critical) failure. We therefore recommend to avoid suppressing False Positives, and either
leave them visible for the developer as warning signs, or even better, rewrite the code in a prover-friendly manner following the tips in Section~\ref{sec:recommendations}.

\textbf{Mistake 2: Inconsistent Contracts.} Function contracts act as
barriers for propagating proof results (besides inlined
functions), that is, the result of a VC in one subprogram cannot affect the result of another in a different subprogram. However, these barriers can be broken when function
contracts are inconsistent, producing False Negatives by our definition. One way to obtain inconsistent contracts, is writing a postcondition which itself contains a failing VC (line 2):
\begin{lstlisting}[name=postvc]
function f1 (X : Integer) return Integer 
  with Post => f1'Result = X + 1 is -- overflow check might fail
begin 
  return X;
end f1;

procedure Caller is
   X : Integer := Integer'Last;
begin
   X := X + 1; -- overflow check proved.
   X := f1(X);
end Caller;
\end{lstlisting}
Clearly, an overflow must happen at line 10, resulting in
an exception. The analyzer, however, proves absence of overflows in
\lstinline$Caller$. The reason is that in the Why3 backend, the
postcondition of \lstinline$f1$ is used as an axiom in the analysis of
\lstinline{Caller}. The resulting theory for \lstinline{Caller} is an
inconsistent axiom set, from which (\emph{principle of explosion})
anything can be proven, including that false VCs are true. In such
circumstances, the solver may
also produce a \emph{spurious} counterexample. 

In the example above, the developer gets a warning for the
inconsistent postcondition and can correct for it, thus keep barriers
intact and ensure that the proofs in the caller are not
influenced. However, if we change line 4 to \lstinline{return X+1}, then the
failing VC is now indicated in the body of \lstinline{f1}, and -- since the
proofs build on each other -- the postcondition is verified and a
defect easily missed. Therefore, failing VCs within callees may also
refute proofs in the caller (in contrast to execution semantics) and
have to be taken into account. Indeed, the textual report of GNATprove (with flag
\texttt{--assumptions}) indicates that AoRTE in \lstinline{Caller} depends on both
the body and the postcondition of \lstinline{f1}, and therefore the reports have to be studied with great care to judge the verification output.
Finally, note that the same principle applies for assertions and loop invariants.

\textbf{Mistake 3: Forgetting the RTS.} Despite proven AoRTE, one
procedure which rotates the frame of reference of the gyroscope
measurements was sporadically triggering an exception after a
floating-point multiplication. The situation was eventually captured
in the debugger as follows:
\begin{lstlisting}[name=subnormal]
-- angle = 0.00429, vector (Z) = -2.023e-38
result(Y) := Sin (angle) * vector(Z);
-- result(Y) = -8.68468736e-41 => Exception
\end{lstlisting}
Variable \lstinline$result$ was holding a \emph{subnormal} floating-point number,
roughly speaking, an ``underflow''. GNATprove models floating-point
computations according to IEEE-754, which requires support for
subnormals on the target processor. Our processor's FPU indeed
implements subnormals, but the RTS, part of which describes
floating-point capabilities of the target processor, was incorrectly
indicating the opposite\footnote{This also has been fixed in recent
  versions of the embedded ARM RTS.}. As a result, the language-defined
float validity check occasionally failed (in our case when the glider
was resting level and motionless at the ground for a longer period of
time). 
  Therefore, the RTS must be carefully configured and checked manually
  for discrepancies, otherwise proofs can be refuted since static
  analysis works with an incorrect premise.

\textbf{Mistake 4: Bad Patterns.} 
\emph{Saturation} may seem like an effective workaround to ensure
overflows, index checks and so on cannot fail, but it usually hides
bigger flaws. Consider the following example, also from the GPS protocol parser:
\begin{lstlisting}[name=saturate]
subtype Lat_Type is Angle_Type range -90.0 * Degree .. 90.0 * Degree;
Lat : Lat_Type := Dim_Type (toInt32 (data_rx(28..31))) * 1.0e-7 * Degree;
\end{lstlisting}
The four raw bytes in \lstinline$data_rx$ come from the GPS device and represent a scaled float, which could in
principle carry a value exceeding the latitude range of $[-90,90]$
Degree. To protect against this sort of error, it is tempting to
implement a function (even a generic) of the form
\lstinline$if X > Lat_Type'Last then X := Lat_Type'Last else...$ that limits the value to the available range, and apply it to all
places where checks could be failing. However, we found that almost every
case where saturation was applied, was masking a
boundary case that needs to be addressed. In this example, we needed
handling for a GPS that yields faulty values. 
In general, such cases usually indicate a missing software requirement.


\subsection{Design Limitations}
We now describe some cases where the current version of the SPARK~2014
\emph{language} -- not the static analysis tool -- imposes limitations.


\textbf{Access types} (pointers) are forbidden in SPARK, however, low-level
drivers heavily rely on them. One workaround is to hide those types in
a package body that is not in SPARK, and only provide a SPARK specification. 
Naturally, the body cannot be verified, but at least its subprograms 
can be called from SPARK subprograms. Sometimes it
is not possible to hide access types, in particular when packages use
them as interface between each other. This is the case for our SD card
driver, which is interfaced by an implementation of the FAT filesystem through access types. Both
are separate packages, but the former one exports restricted types and
access types which are used by the FAT package, thus requiring that wide parts of the FAT package 
are written in Ada instead of SPARK.

As a consequence, access types are sometimes demanding
to form larger monolithic packages, here to combine SD card driver and FAT filesystem
into one (possibly nested) package.

\textbf{Polymorphism} is available in SPARK, but its use is
limited as a result of the access type restriction. Our message
queue between flight-critical and mission-critical task was planned to
hold messages of a polymorphic type. However, without access types the
only option to handover messages would be to take a deep copy and
store it in the queue. However, the queue itself 
is realized with an array and can hold only objects of the
same type. This means a copy would also be an upcast to the base
type. This, in turn, would loose the components specific to the derived
type, and therefore render polymorphism useless. As a workaround, we
used mutable variant records.

\textbf{Interfaces.} Closely related to polymorphism, we intended to implement sensors as polymorphic types. That is, specify an abstract sensor interface that must be overridden by each sensor implementation. 
Towards that, we declared an abstract tagged type 
with abstract primitive methods denoting the interface that a specific sensor must implement. 
However, when we override the method for a sensor implementation, such as the IMU, SPARK requires specifying the global dependencies of the overriding IMU implementation as class-wide global dependencies of the abstract 
method (SPARK RM 6.1.6). This happens even without an explicit \texttt{Global} aspect.
%
As workaround, we decided to avoid polymorphism and used simple inheritance without overriding methods.

\textbf{Dimensioned Types.}
Using the GNAT dimensionality checking system in SPARK, had revealed two missing features. Firstly, in the current stable version of the GNAT compiler, it is not possible to specify general operations on dimensioned types that are resolved to specific dimensions during compilation.
For example, we could not write a generic time integrator function for the PID controller that multiplies any dimensioned type with a time value and returns the corresponding unit type. Therefore, we reverted to dimensionless and unconstrained floats within the generic PID controller implementation. 
Secondly, it is not possible to declare vectors and matrices with mixed subtypes, which would be necessary to retain the dimensionality information throughout vector calculations (e.g., in the Kalman Filter). As a consequence, we either have split vectors into their components, or reverted to dimensionless and unconstrained floats. 
As a result of these workarounds, 
numerous overflow checks related to PID control and Kalman Filter could not be proven (which explains more than \SI{70}{\percent} of our failed floating-point VCs). 

\subsection{Solver Weaknesses}
We now summarize some frequent problems introduced by the current
state of the tooling. 

The \texttt{'Position} attribute of a record allows evaluating the
position of a component in that record.
However, GNATprove has no precise information
about this position, and therefore proofs building on that might fail.

Another feature that is used in driver code, are
\emph{unions}, which provide different views on the
same data. GNATprove does
not know about the overlay and may generate False Positives for
initialization, as well as for proofs which build on the relation
between views.


We had several False Positives related to possibly uninitialized
variables.  SPARK follows a strict data initialization policy. Every
(strict) output of a subprogram must be initialized. In the current version,
GNATprove only considers initialization of arrays as complete when done in a single
statement. This generates warnings when an array is initialized in
multiple steps, e.g., through loops, which we have suppressed.



\section{Results}
\label{sec:results}

\begin{figure}[hbtp]\vspace*{-4mm}
    \begin{subfigure}[t]{0.48\textwidth}
        \includegraphics[width=\textwidth]{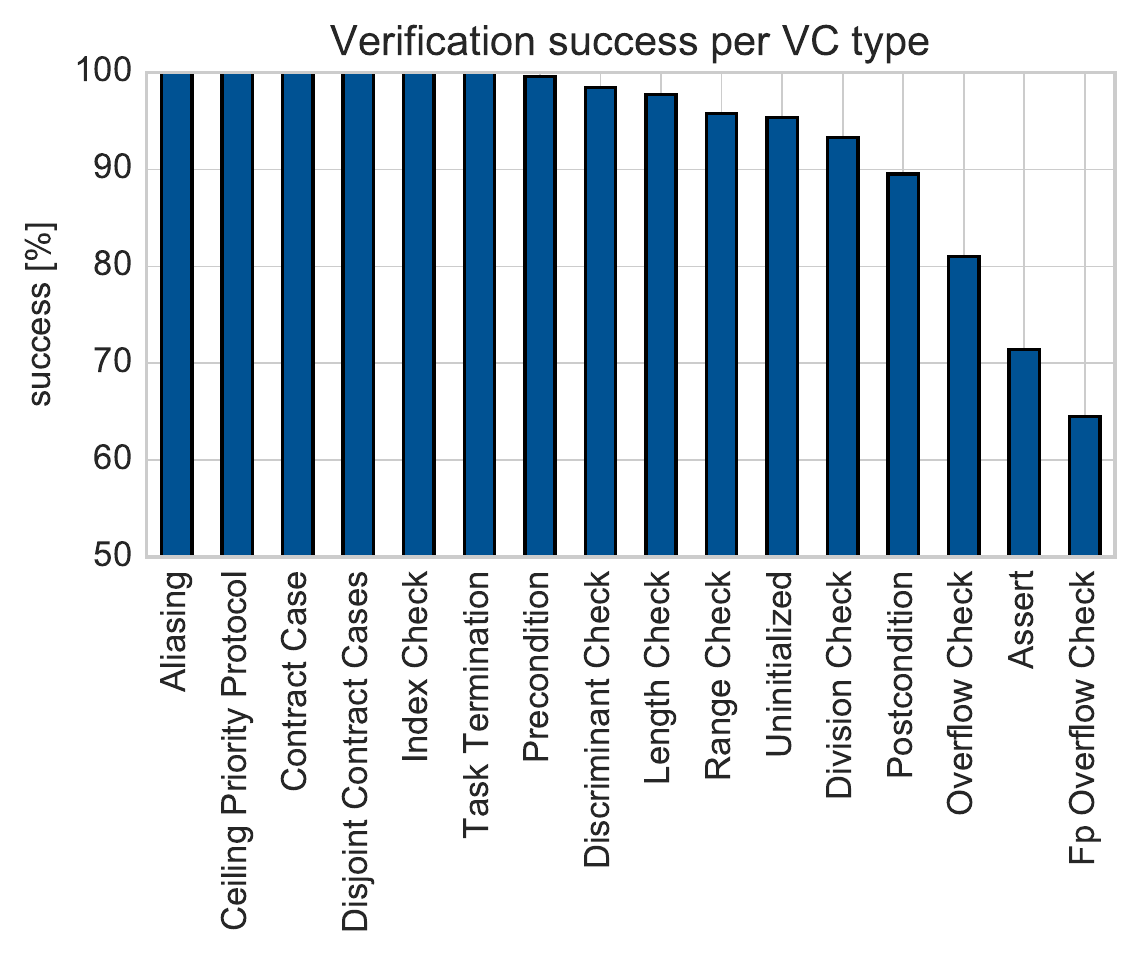}
    \end{subfigure}
    ~
    \begin{subfigure}[t]{0.48\textwidth}
      \includegraphics[width=\textwidth]{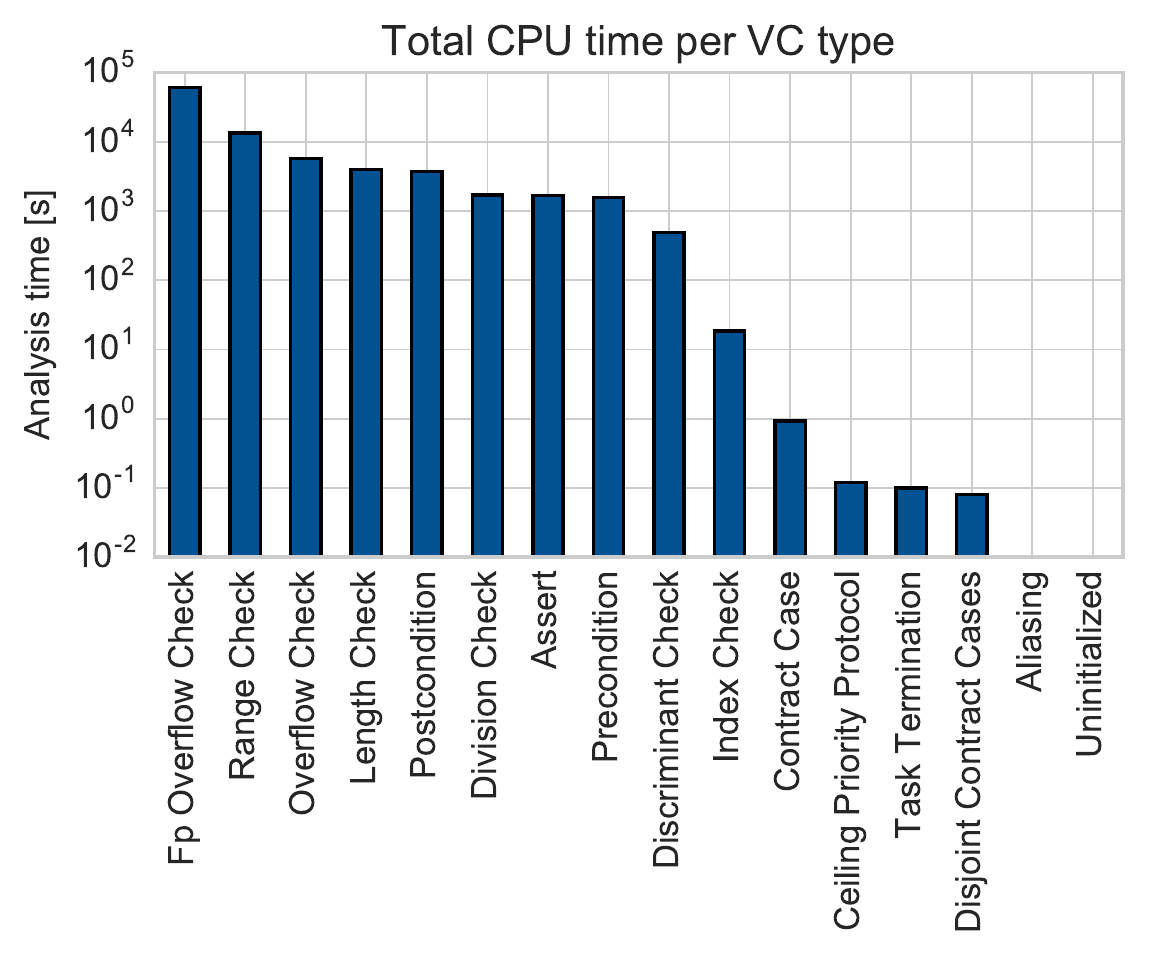}
    \end{subfigure}
    \caption{Statistics on Verification Conditions (VCs) by type.
\vspace*{-3mm}}\label{fig:stats}
  \end{figure}

In general, verification of SPARK~2014 programs is accessible and mostly
automatic.  Figure~\ref{fig:stats} shows
the results of our launch release. As it can be seen, we could not
prove all properties during the time of this project (three months). The non-proven
checks have largely been identified as ``fixable'', following our design recommendations given below.

The complexity of our flight stack and verification progress are
summarized in Table~\ref{tab:complexity}. It can be seen that our
focus on the application part is reflected in the SPARK coverage that we
have achieved (\SI{82}{\percent} of all bodies in SPARK, and even
\SI{99}{\percent} of all specifications), but also that considerably more work has to
be done for the BSP (currently only verified by testing). In
particular, the HAL (off-chip device drivers, bus configuration, etc.)
is the largest part and thus needs a higher SPARK coverage.
However, we should add that \SI{43}{\percent} of the HAL is consisting of
specifications generated from CMSIS-SVD files, which do not contain any
subprograms, but only definitions of peripheral addresses and record definitions to access
them, and therefore mostly cannot be covered in SPARK. Last
but not least, a completely verified RTS would be desirable, as well.

\bgroup
\setlength\tabcolsep{.5em}
\begin{table}[htbp]\vspace*{-9mm}
\begin{minipage}{\linewidth}
  \centering
  \caption{Metrics and verification statistics of our Flight Stack.}
  \small
  \begin{tabular}{lrrrr}
    \toprule
    &               & \multicolumn{2}{c}{\scriptsize{Board Support Package}} & \\
                         \cmidrule{3-4}
    Metric                      & {Application} & {HAL}         & RTS          & {All}\\
    \midrule
    lines of code (GNATmetric)  & 6,750         & 32,903        & 15,769       & 55,422   \\
    number of packages          & 49            & 100           & 121          & 270      \\
    cyclomatic complexity       & 2.03          & 2.67          & 2.64         & 2.53      \\
    SPARK body/spec             & 81.9/99.4\,\% & 15.5/23.5\,\% & 8.6/11.8\,\% & 30.0/38.5\,\%  \\ 
    \midrule
    number of VCs               & 3,214         & 765      & 2       & 3,981      \\
    VCs proven                  & 88.1\,\%      & 92.5\,\% & 100\,\% & 88.8\,\%   \\
    analysis time\footnote{Intel Xeon E5-2680 Octa-Core with \SI{16}{GB} RAM, timeout=\SI{120}{\second}, steps=inf.}  & --        & --        & -- & \SI{19}{\minute} \\
    \bottomrule
  \end{tabular}\label{tab:complexity}
\end{minipage}
\vspace*{-4mm}
\end{table}
\egroup

\textbf{Floats are expensive}. Statistically, we have spent most
  of the analysis time (\SI{65}{\percent}) for proving absence of floating-point
  overflows, although these amount to only \SI{21}{\percent} of all VCs. This is
  because discharging such VCs is in average one
  magnitude slower than discharging most other VC types. In particular, one has
  to allow a high step limit (roughly the number of
  decisions a solver may take, e.g., deciding on a literal) and a
  high timeout. Note that at some point an increase of either of them does not improve 
  the result anymore.

\textbf{Multi-Threading} could be proven to follow our goals. By using the Ravenscar RTS, 
our goals related to deadlock, priority inversion and blocking, hold true by design.
Several
race conditions and non-thread-safe subprograms have been identified by GNATprove, which otherwise would have refuted task separation. To ensure that
termination of low-criticality tasks cannot terminate the flight-critical task, we
provided a custom implementation for GNAT's last chance handler 
(outside of the SPARK language and therefore not being analyzed) which
reads the priority of the failing task 
and acts accordingly: If the priority is lower than
that of the flight-critical task (i.e., the mission-critical
task had an exception), then we prevent a system reset by sending the
low-priority task into an infinite null loop (thus keeping it busy
executing \texttt{nop}s, and keeping the flight-critical task alive). 
If the flight-critical task is failing, then our handler allows a system reset. 
Multi-threading is therefore easy to
implement, poses no verification problems, and can effectively
separate tasks by their criticality.

\textbf{High-Level Behavioral Contracts} related to the homing functionality could
be expressed and proven with the help of ghost functions, although this is beyond the main purpose of SPARK contracts. For example, we could prove the overall behavior in case of loosing the GPS fix, or missing home coordinates.


\subsection{Design Recommendations}\label{sec:recommendations}
The following constructs and strategies have been found amenable to verification:\vspace{-1mm}
\begin{enumerate}
\item Split long expressions into multiple statements $\rightarrow$ discharges more VCs. 
\item Limit ranges of data types, especially floats $\rightarrow$ better analysis of overflows.
\item Avoid saturation $\rightarrow$ uncovers missing error handling and requirements.
\item Avoid interfaces $\rightarrow$ annotations for data flows break concept of abstraction.
\item Emulate polymorphic objects that must be copied with mutable variant records.
\item Separation of tasks by criticality using a custom last chance handler $\rightarrow$ abnormal termination of a low-criticality task does not cause termination of high-criticality tasks.
\end{enumerate}


\section{Related Work}
\label{sec:related-work}

Only a small number of experience reports about SPARK~2014 have been
published before. A look back at (old) SPARK's history
and its success, as well as an initial picture of SPARK~2014 is given
by Chapman and Schanda in~\cite{Chapman2014}. We can report that the
mentioned difficulties with floating-point numbers are solved in SPARK~2014, 
and that the goal to make verification more accessible, has been reached.  A small case study with SPARK~2014
is presented in~\cite{Trojanek2014}, but at that point multi-threading
(Ravenscar) was not yet supported, and floating point numbers have
been skipped in the proof. We can add to the conclusion given there,
that both are easily verified in ``real-world'' code, although
float proofs require more (computational and mental) effort.  Larger
case studies are summarized by Dross et al. in \cite{Dross2014}, with
whom we share the opinion of minor usability issues, and that some
small amount of developer training is required.  Finally, SPARK~2014
with Ravenscar has recently been announced to be used in the Lunar
IceCube~\cite{Brandon2016} satellite, a successor of the successful CubeSat project that was implemented in SPARK 2005. It will be a message-centric software, conceptually similar to NASA's cFE/CFS, but 
fully verified and striving to become an open source platform for spacecraft software.
In contrast to all the above publications, this paper is not focused on the application or case studies, but  pointing out typical sources of errors in SPARK programs, which a developer has to
know in order to get correct verification results.

\section{Conclusion\label{sec:conclusion}}
Although the verification of SPARK~2014 programs is very close to
execution semantics and therefore mostly intuitive, we believe that
developers still need some basic training to avoid common mistakes as
described in this paper, which otherwise could lead to a false confidence in
the software being developed. 
Overall, the language forces developers to address boundary
cases 
of a system explicitly, which eventually helps understanding the
system better, and usually reveals missing requirements for
boundary cases. As a downside, SPARK~2014 programs are often longer than
(approximately) equivalent Ada programs, since in the latter case a
general exception handler can be installed to handle all pathological
cases at once, without differentiating them. Furthermore, static analysis is ready
to replace unit tests, but integration tests have still been found
necessary. 

Regarding the shortcomings of the GNAT dimensionality system, we
can report that as a consequence of our experiments, a solution for
generic operations on dimensioned has been found and will be part of future GNAT releases.

Our remaining criticism to SPARK~2014 and its tools is as follows:
next to some minor tooling enhancements to avoid the mistakes
mentioned earlier and adding some more knowledge to the analyzer, it
is necessary to support object-oriented features in a better
way. All in all, SPARK~2014 raises the bar for formal
verification and its tools, but developers still have to be aware of
limitations.


\section*{Acknowledgements}
Thanks to the SPARK~2014 team of AdaCore for their guidance and insights.

{
 \scriptsize
 \bibliographystyle{splncs03}
 \bibliography{literature}
}

\end{document}